\documentclass[12pt]{article}
\usepackage{amsmath}
\usepackage{amssymb}
\usepackage{graphicx}
\usepackage{epsfig}
\usepackage[all,knot,poly,dvips]{xy}

\begin{document}

\begin{center}
{ \large \bf
Post Quantum Cryptography\\ 
from Mutant Prime Knots }
\end{center}

%\begin{center}
%{\large \today}
%\end{center}

\vspace{12pt}

\begin{center}
{\large {\em Annalisa Marzuoli}$^{\,(1)}$ and {\em Giandomenico Palumbo}$^{\,(2)}$}
\end{center}

\vspace{6pt}

\begin{center}
Dipartimento di Fisica Nucleare e Teorica,
Universit\`a degli Studi di Pavia and
Istituto Nazionale di Fisica Nucleare, Sezione di Pavia\\
via A. Bassi 6, 27100 Pavia (Italy)\\
$^{\,(1)}$ E-mail: annalisa.marzuoli@pv.infn.it \\
$^{\,(2)}$ E-mail: giandomenico.palumbo@pv.infn.it 
\end{center}

\vspace{6pt}

\begin{abstract}
By resorting to basic features of topological 
knot theory we propose a (classical) cryptographic protocol based
on the `difficulty' of decomposing
complex knots generated as connected sums of prime knots and their mutants.
The scheme combines an asymmetric public key protocol
with symmetric private ones and is intrinsecally
secure against quantum eavesdropper attacks.  
\end{abstract}

\vspace{4 cm}

\noindent 
{\bf PACS2008}:\\ 
89.70.-a (Information and communication theory)\\
02.10.kn (Knot theory)\\
03.67Dd (Quantum Cryptography and communication security)

\vspace{6pt}

57M27 (Invariants of knots and 3-manifolds)\\
68Q17 (Computational difficulty of problems)

\vfill
\newpage

\section{Introduction}

Knots and links (collections of knotted circles),
beside being fascinating mathematical objects, 
are encoded in the modeling
of a number of physical, chemical and biological systems.
In particular it was in the late 1980 that knot theory was recognized to have
a deep, unexpected interaction with quantum field theory \cite{Wit}. 
In earlier periods of the history
of science, geometry and physics interacted very strongly at the `classical' level
(as in Einstein's General Relativity theory), but the main feature of this new,
`quantum' connection is the fact that geometry is involved in a global and not purely local
way, {\em i.e.} only `topological' features do matter. 
Over the years mathematicians have proposed a number of `knot
invariants' aimed to classify systematically all possible knots. 
Most of these invariants
are polynomial expressions (in one or two variables) with coefficients in the relative integers. 
It was Vaughan Jones in \cite{Jo85} who discovered the most famous polynomial invariant, 
the Jones invariant, and solved the Tait's conjectures for alternating knots. 
In the seminal paper by Edward Witten \cite{Wit}, the Jones polynomial was actually recognized to be 
associated with the vacuum expectation value 
of a `Wilson loop operator' in a quantum Chern--Simons theory (see the rewiews
\cite{KaGoRa}, \cite{Boi} for  comprehensive  accounts on these topics).

Seemly far from the above remarks, the search for new algorithmic problems and techniques 
which should improve `quantum'  with respect to classical computation is getting more and more 
challenging in the last decade. 
Most quantum algorithms are based on the standard quantum circuit model \cite{NiCh},
and are designed to solve problems which are essentially number theoretic such as the
Shor's algorithm \cite{Sho} (see \cite{EkMa} for a general review on the basics
of quantum algorithms). However, other
types of problems, typically classified in the field of enumerative combinatorics 
and ubiquitous in many areas of mathematics and physics, 
share the feature to be `intractable' in the framework of classical information theory. 
In particular the evaluation of the Jones polynomial has been shown to be 
$\mathbf{\# P}$--hard, namely computationally intractable in a very strong sense \cite{JaVeWe}.
In this  perspective, 
efficient quantum algorithms for computing approximately 
knot invariants (of the Jones' type or extensions of it)
have been successfully addressed in the last few years
\cite{AhJoLa},  \cite{GaMaRa1,GaMaRa2,GaMaRa3}, \cite{WoYa}
and indeed such problem has been recognized to be  `universal'
in the quantum complexity class
{\bf BQP}
({\bf B}ounded error {\bf Q}uantum {\bf P}olynomial), namely the hardest problem that a quantum computer can 
efficiently handle \cite{BoFrLo}.

Notwithstanding the improvements outlined above both
in  field--theoretic settings  and in quantum complexity
theory, the basic unsolved problem in topological knot theory
still remains the `recognition problem'. Namely, given
two knots, how can we check if they are `equivalent'
(in the sense to be formalized in the next section).
Invariants of (oriented) knots might be useful to this task,
but there exist particular classes of knots
--the `mutants' of a given knot-- that
cannot be distinguished {\em in principle} 
since by  definition all of them possess the same
Jones' type invariants, a result derived 
by resorting to standard tools in combinatorial topology
(see {\em e.g.} \cite{Lic}) but recognizable also
in the field--theoretic framework
as a property of expectation values 
of Wilson loop operators  
\cite{RaGoKa}.\\

As is well known,
group--based cryptography has became in the last few years
a very fruitful branch of cryptoanalysis  \cite{BlCiMu}, \cite{Gon}.
In particular, the key--agreement protocol
proposed in \cite{AnAnGo} can be implemented using 
the braid group $\mathbf{B}_n$ (a non--Abelian group on
(n-1) generators that can be associated to geometric configurations
of n interlaced strands whose endpoints are fixed on two
parallel straight lines in the plane).
Knots and braids are indeed 
closely interconnected since we can get a (multi--component) knot
by `closing' up an open braid, and a number of interesting algorithmic problems 
related to this group can be addressed \cite{BiBr}. 
Roughly speaking, a braid--group--based cryptographic protocol  
relies on the existence of an `easy' problem
(recognize whether two braids $W$ and $W'$, expressed algebraically
in terms of generators of the braid group, are the same element)
and an `intractable' one (recognize whether two words $W_1$ and $W_2$
are conjugate to each other, namely if there exists a
$W'$ for which $W_2=W'W_1(W')^{-1}$).
As reviewed in \cite{BlCiMu}, basic ingredients for
implementing secure cryptosystem are the computational
time required to execute the protocol, 
the number of bits that are to be exchanged between
Alice and Bob, the number of passes (exchange of information), the
sizes of keys and the sizes of system parameters.
However modern security is often much more demanding,
so that at  present braid--group--based protocols \cite{AnAnGo,Deh}
do not seem safe from eavesdropper attacks.\\
 
The 
theoretically secure protocol we propose in this paper
is framed within  topological knot theory
and the basic ingredients are  `prime' knots depicted  in a standardized
manner in Knot Tables currently available on the web. The scheme 
relies on the `easy' problem of associating with prime knots
in Knot Tables their Dowker--Thistlethwaite codes, 
numerical sequences which
are different for inequivalent knots. Then we
resort to the `difficulty' of factorizing, so to speak,
complex knots generated by composing prime knots and their mutants.
The scheme resorts to purely classical cryptographic tools,
combining an asymmetric public key protocol
with symmetric private ones.

The adjective  `post quantum'  in the title 
comes about {\em a posteriori} in  light of the fact that  
most currently popular public--key cryptosystems 
rely on the integer factorization problem or discrete logarithm problem
(arising {\em e.g.} in the framework of cyclic group--based protocols),
both of which would be easily solvable 
on large enough quantum computers using Shor's algorithm. Our protocol is not based 
on the quoted two problems, neither seems  reducible to them, and
thus the standard meaning of post--quantum --secure against `quantum' attacks--
can be taken for granted until someone will be able to prove the converse.
In a somehow extended sense, and according to the
remarks made above on quantum algorithms for 
computing knot invariants,  an attack based on (quantum)
calculations of such polynomials would fail 
in view of the presence of mutants, not detectable 
even by a quantum computer.

In section 2 we review in brief some basic notions in topological
knot theory, while in section 3 the cryptographic protocol
is presented. A few more comments and conclusions are collected in section 4.

%%%%%%%%%%%%%%%%%%%%%%%%%%%
%%%%%%%%%%%%%%%%%%%%%%%%%%%%%%%%%%
\section{Overview of Topological Knot Theory\\
and coding of knot diagrams}
%%%%%%%%%%%%%%%%%%%%%%%%%%%%%%%%
%%%%%%%%%%%%%%%%%%%%%%%%%%%%%%%%%%%%%

A {\em knot} $K$ is defined as a continuous embedding of the circle $S^1$ (the $1$--dimensional 
sphere) into the Euclidean $3$--space $\mathbb{R}^3$ or, equivalently, into the $3$--sphere
$S^3 \doteq \mathbb{R}^3 \cup \{\infty\}$. A {\em link} $L$ is the embedding of the disjoint 
union of $M$ circles, $\cup_{m=1}^{M}\,(S^1)_m$ into $\mathbb{R}^3$ or $S^3$, namely a finite 
collection of knots.  Since each circle can be naturally endowed with an orientation, 
we can introduce naturally {\em oriented} knots (links).

Referring for simplicity to the unoriented case, two knots $K_1$ and $K_2$ are said to be 
{\em equivalent}, $K_1 \sim K_2$, if and only if they are (ambient) isotopic. An isotopy
can be thought of as a continuous deformation of the shape of, say, $\,K_2 \subset \mathbb{R}^ 3$
which makes $K_2$ identical to $K_1$ without cutting and gluing back the `closed string' 
$K_2$.

The {\em planar diagram}, or simply the {\em diagram}, of a knot $K$ is the projection of $K$
on a plane $\mathbb{R}^2 \subset \mathbb{R}^3$, in such a way that no point belongs to
the projection of three segments, namely the singular points in the diagram are only
transverse double points. Such a projection, together with `over' and `under' information
at the crossing points --depicted in figures by breaks in the under--passing segments--
is denoted by $D(K)$. In what follows we 
shall  identify the symbols $K$  with $D(K)$, although we can 
obviously associate with a same knot an infinity of planar diagrams.

%%%%%%%%%%%%%%%%%%%%%
%%%%%%%%%%%%%%%%%%%%%%%
The number of crossings of a knot (diagram) is clearly a good indicator of the `complexity'
of the knot and indeed Tait in  late 1800 initiated a program aimed to classifying systematically 
knots in terms of the number of crossings. 
In Knots Tables (see \cite{HoThWe} and the \textit{Knot Atlas} on Wikipedia)
there appear diagrams of unoriented `prime' knots listed by increasing crossing numbers
as $\digamma_N$, where $\digamma$ is the number of crossings and
$N= 1,2,\dots$ enumerates in a conventional way the (standard projections of)
knots with the same  $\digamma$. 
The (unique) `unknot' or trivial knot $K_{\bigcirc}$
has standard projection given by the  circle, {\em i.e.} $\digamma_N 
(K_{\bigcirc})=0$
with $N=1$.
A prime knot is defined
as a non--trivial knot which 
cannot be decomposed into two (or more) non--trivial knots. 
Decomposition is in turn the inverse of the topological
operation of composition of knot diagrams.
More precisely, given two knot diagrams $K_1$ and $K_2$, it is possible to
draw a new knot  by removing a small segment from each knot and
then joining  the four endpoints by two new arcs.
The resulting diagram is the {\em connected sum} of $K_1$, $K_2$,
denoted by $K_1$ $\#$ $K_2$. As shown below, starting for instance 
from the diagrams of the trefoil knot
$K_1$ (configuration $3_1$ in Knot Tables)
and its mirror image $K_2$, 
their connected sum turns out to be the so--called `square' knot,
the six--crossings configuration listed as $6_2$.
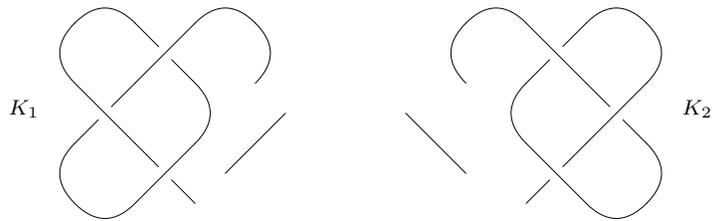
\[
\renewcommand{\labelstyle}{\scriptstyle}
\begin{xy}
0;/r8mm/:
,{\xcapv-|{\,}}
, +(0,1) ,
{\xcaph|{\,}
\xunderh|{\,}%
\xcaph|{\,}
\xcapv|{\,}}
, -(3,0),{\xoverh|{K_1\;\;\;\;\;\;}}
,-(1,1),{\xcapv-|{\,}\xcaph-|{\,}}
, +(0,1),{\xunderh|{\,}}
, +(0,1),{\xcapv|{\,}}
, +(0.5,-0.5),{\sbendh @(0)}
, +(2,0),{\sbendv @(0)}
, +(1,1.5)  , 
{\xcapv-|{\,} 
\xoverh|{\,}}
,+ (0,-1),{\xcaph-|{\,}}
,+(0,1), {\xcapv|{\,}}
, +(-1,2) ,
\xunderh|{\;\;\;\;\;\;K_2} 
, +(0,1) \xcapv|{\,}
, +(-1,1) \xcaph|{\,} 
,-(2,0) \xoverh|{\,}
, +(-2,0)\xcaph|{\,}
, +(-1,0) \xcapv-|{\,}
\end{xy}
\]
\[
\renewcommand{\labelstyle}{\scriptstyle}
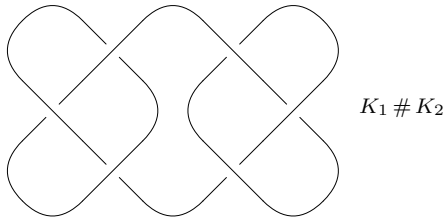
\begin{xy}
0;/r8mm/:
,{\xcapv-|{\,}}
, +(0,1) ,
{\xcaph|{\,}
\xunderh|{\,}%
%\xcaph|{d}
%\xcapv|{e}
}
, -(2,1),{\xoverh|{\,}}
,-(1,1),{\xcapv-|{\,}\xcaph-|{\,}}
, +(0,1),{\xunderh|{\,}}
, +(0,1),{\xcapv|{\,}}
%, +(0.5,-0.5),{\sbendh @(0)}
%, +(2,0),{\sbendv @(0)}
, +(0,2) , \xcaph|{\,}
, +(0,-1)  , 
{\xcapv-|{\,} 
\xoverh|{\,}}
, +(-2,-1) \xcaph-|{\,}
,+ (1,0),{\xcaph-|{\,}}
,+(0,1), {\xcapv|{\,}}
, +(-1,2) ,
\xunderh|{\;\;\;\;\;\;K_1 \, \mathbf{\#} \,K_2} 
, +(0,1) \xcapv|{\,}
, +(-1,1) \xcaph|{\,} 
,-(2,0) \xoverh|{\,}
%, +(-2,0)\xcaph|{l}
%, +(-1,0) \xcapv-|{m}
\end{xy}
\]
The connected sum of knot diagrams (well defined for oriented knots)
is commutative and associative
and has an identity element given by the trivial knot $K_{\bigcirc}$,
namely $K_{\bigcirc}$ $\#$ $K$ $= K$ for each $K$.
Remarkably, to each diagram representing a composite knot  it is possible to associate a
decomposition into prime knots which is unique \cite{Sch} --up to ordering of summands.
The (minimal) crossing number used for building up Knot Tables is
the first example of a numerical knot `invariant' since
it depends only on the ambient isotopy class 
of the knot. Switching to knot (link) diagrams, 
it can be proved that a knot invariant is a
quantity (a number or a polynomial, see below) which does not change
under applications to the diagrams of
finite sequences of  the so--called Reidemeister moves
(we leave aside this issue and refer to the
classic books \cite{Rol, Lic, Kau, Ada} 
also as general references
on knot invariants).
It is not difficult to recognize  that polynomial
invariants can take the same value on  inequivalent knots,
and it is the biggest open problem in knot theory to
establish a `complete' set of invariants able to distinguish
(and thus classify) all equivalence classes of knots.
Most famous polynomial invariants of knots, 
such as  Alexander polynomial,  Jones polynomial \cite{Jo85}
and its extensions \cite{ReTu}
(in one formal variable) as well as  HOMFLY \cite{HOMFLY}
and Kauffman \cite{Kau} polynomials (in two variables)
are able to distinguish particular sub--classes or types of  knots. 
Actually, even resorting to all of them,
there exists quite a large number of examples
(with relatively small crossing numbers)
in which indistinguishable diagrams still remain.
In particular, neither  Jones, Kauffman and HOMFLY polynomials, nor
more general invariants such as Reshetikhin--Turaev ones, are  sufficient to
distinguish {\em any} knot $K$ from  its  mutations $K'$ \cite{Lic,RaGoKa}. 

To explain what is a `mutant' knot we introduce first
a `tangle' notation for dealing with knot diagrams. 
A tangle is defined as a region of the planar diagram of an oriented 
or unoriented knot
bounded  by a circle (not belonging to the diagram)
such that the knot strands cross
the circle exactly four times. Thus
any knot can be always presented by resorting to
(at least) two tangles, say $\mathfrak{S}$ and $\mathfrak{R}$,
joined by 2+2 strands (this shorthand
graphical notation for a single knot
should not be confused
with the operation of connected sum on
knot diagrams).
 
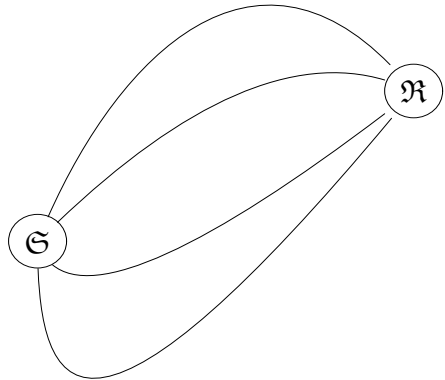
\[
\xy
*+[o]+[F]{\mathfrak{S}};
(50,20) 
*+[o]+[F]{\mathfrak{R}}="R"
, "R" **\crv{(30,30)}
, "R" **\crv{(10,25)&(30,45)}
, "R" **\crv{(5,-15)}
, "R" **\crv{(0,-20)&(5,-35)}
\endxy
\]

Starting from  a tangle presentation of an {\em oriented} knot $K$, 
a mutant  $K'$ arises  by removing, {\em e.g.}, the tangle 
labeled by $\mathfrak{R}$ (two strands ingoing
and two outgoing) and replacing it
with a tangle $\mathfrak{R}'$ obtained
by rotating $\mathfrak{R}$
(and reversing orientation of some strands
if necessary). Admissible rotations 
are depicted below: the inner content of the tangle undergos 
$\pi$--rotations with respect to three mutually orthogonal axes
which can be thought as pointing from the central
configuration toward the other three  embedded in a reference 3--space.
Note that only two of these rotations are independent,
but of course the process of mutation can be carried out at will
on different subsets of a same knot diagram including
at least one crossing.

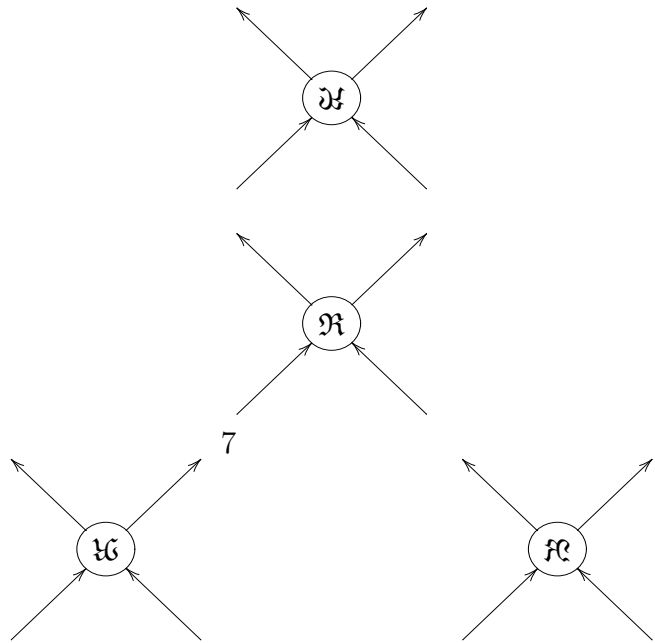
\[
\begin{xy}
0* {\xymatrix{
 \,  & \, & \,  \\
& *+[o]+[F]{\scalebox{1}[-1]{$\mathfrak{R}$}} \ar[ul] \ar[ur] & \, \\
  \, \ar[ur] & \, & \, \ar[ul]
}} ,
<0cm,-1.5cm>* {\xymatrix{
 \,  & \, & \,  \\
& *+[o]+[F]{\mathfrak{R}}  \ar[ul] \ar[ur] & \, \\
  \, \ar[ur] & \, & \, \ar[ul]}}
,
<1.5cm,-3cm>*
{\xymatrix{
 \,  & \, & \,  \\
& *+[o]+[F]{\reflectbox{$\mathfrak{R}$}} \ar[ul] \ar[ur] & \, \\
  \, \ar[ur] & \, & \, \ar[ul]
}} ,
<-1.5cm,-3cm>*
{\xymatrix{
 \,  & \, & \,  \\
& *+[o]+[F]{\rotatebox{180}{$\mathfrak{R}$}} \ar[ul] \ar[ur] & \, \\
  \, \ar[ur] & \, & \, \ar[ul]
}}
\end{xy}
\]

\vfill
\newpage
In view of applications in cryptography
we conclude this section by introducing 
Dowker--Thistlethwaite (DT) notation (or code) for oriented knots.
This allows us to associated to each 
planar diagram its (minimal) DT sequence (actually
a string of relative integers) from which it is possible 
to reconstruct (almost) uniquely the knot.
Consider as an example an oriented alternating knot with n
crossings (namely a diagram with an alternating sequence 
of over and under--crossings) and start 
labeling an arbitrary crossing with  1. Once fixed an orientation,
go down the strand to the next crossing and denote it by 2.
Continue around the knot  until each
crossing has been numbered twice. Then each crossing 
is decorated with a pair of even/odd positive numbers, 
running from 1 to 2n, as shown below for the knot
$5_1$.
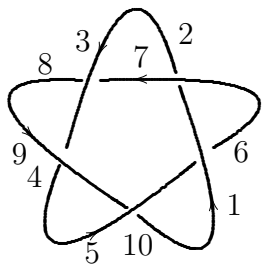
\[\xy
(6,9)*{}="1";
(-8.5,-1)*{}="2";
"1";"2"**\crv{~*=<.5pt>{.}
(0,30)}?(.75)*\dir{>}+(-2,2)*{3}+(-5,-3)*{8};
(-6.5,8)*{}="1";
(-.5,-9)*{}="2";
"1";"2"**\crv{~*=<.5pt>{.}
(-28.5,9.3)}?(.7)*\dir{>}+(-2,-2)*{9}+(2,-3)*{4};
(-9.5,-3.35)*{}="1";
(8.5,-3)*{}="2";
"1";"2"**\crv{~*=<.5pt>{.}
(-17.67,-24.19)}?(.7)*\dir{>}+(-1,-3)*{5}+(6,1)*{10};
(1,-10)*{}="1";
(6.5,7.13)*{}="2";
"1";"2"**\crv{~*=<.5pt>{.}
(17.67,-24.19)}?(.7)*\dir{>}+(3,-1)*{1}+(1,7)*{6};
(11,-1)*{}="1";
(-4,8)*{}="2";
"1";"2"**\crv{~*=<.5pt>{.}
(28.5,9.3)}?(.93)*\dir{>}+(1,3)*{7}+(6,3)*{2};
\endxy\]
For prime alternating knots the notation uniquely defines a single knot 
in case of amphichiral knots or corresponds to a single knot or 
its mirror image in case of chiral ones. 
(Recall that a chiral knot is a knot that is 
not equivalent to its mirror image
while an oriented knot that is equivalent 
to its mirror image is an amphichiral knot).
For generic, non--alternating prime knots 
(which actually appear in tables for crossing numbers greater than
7), the Dowker--Thistlethwaite coding is slightly modified by making 
the sign of the even numbers positive
if the crossing is on the top strand, and 
negative  if it is on the bottom strand. 
Since any Dowker sequence is dependent on both a minimal
projection and  the choice of a starting point, the mapping
between knots and their DT sequences is one--to--many, so it would be
necessary to find a minimal DT sequence for each (composite) knot. Hence
DT codes are to be interpreted as minimal permutations 
of strings of relative integers representing certain knot
diagrams, not carrying significant  topological
information about knots (so that they are useless in any
attempt of knot classification).

Summing up, the assignment of Dowker--Thistlethwaite codes
to prime knots enumerated in currently available Knot Tables
(up to 17 crossings) is essentially unique and the length
of these numerical strings (plus possibly $\pm$ signs)
grows linearly with the crossing number.

\section{Cryptography using knots}
Let us remind some basics facts about
RSA cryptosystem, the most famous protocol
of all times  invented by Rivest, Shamir and Adleman \cite{RSA}
and  based on the concept of `asymmetric' public key.
Imagine that  A (Alice) must send a secret message to B (Bob). 
It would take the following steps:

\begin{itemize}
\item[1.] B generates a public key $\chi$ by resorting to a certain set
of `generators'.

\item [2.] B sends the public key to A. Anyone can see it.
  
\item[3.] A uses the key to encrypt the message $\mathcal{M}$;

\item[4.] A sends the encrypted message $\mathcal{M}^{\chi}$ to B, 
but none can decrypt it.
   
\item[5.] B receives the message $\mathcal{M}^{\chi}$ and, knowing the generators, 
is able to decrypt it.

\end{itemize}

Actually most RSA--type protocols are based on the  computational complexity of 
factorization of prime numbers, 
because the generators are two large prime numbers (p
and q) and the public key  is the product of them (N = pq).
Once given N, decrypting the message needs the knowledge 
of its prime factors, and
this is of course a computationally hard problem. Note however that
public key algorithms are very costly in terms of
computational resources. The time it takes the message to be encoded
and decoded is relatively high and  this is actually
the main drawback of (any)
asymmetric decoding. This problem can be overcome or even solved
by using a symmetric key together with the asymmetric one,
as we are going to illustrate in the following statements defining our
knot--based cryptosystem.\\

A must send a secret message to B and they share
the same finite list of prime knots $K$'s. 
The message $\mathcal{M}$ will be  built by resorting to
a finite sequence of (not necessarily prime) knots $L_{1},...,L_{N}$ as described below   

\begin{itemize}

\item[step I)] Through a standard RSA 
protocol, B sends to A an ordered  sublist of $N$
prime knots (taken from current available Knot Tables) $K_{1},...,K_{N}$,
together with mutation instructions to be applied to each $K_{i}$
(also no mutation on some of them is allowed).
Then  a second list $K'_{1},...,K'_{N}$ is generated by
picking up definite mutations of the original sequence. 
 
\item[step II)]  A takes $K'_{1},...,K'_{N}$ 
 and performs a series of ordered connected sums
$$ L_{1} \# K'_{1},\, L_{2} \# K'_{2},\, \ldots, L_{N} \# K'_{N}
$$
with the knots $L_{1},...,L_{N}$ associated with the message to be sent.\\
These composite  knots are now translated (efficiently) into 
Dowker--Thistlethwaite sequences and sent to B. Obviously 
at this stage everyone has access to these 
strings of relative integers.
   
\item[step III)] B receives the (string of) composite knots. 
Since he knows the DT sub--codes
for the prime knots of the shared list, 
he can decompose the composite knots, thus
obtaining the DT code for every $L_{i}$. 
Then the planar diagrams of 
$L_{1},...,L_{N}$ can be uniquely recovered. 
\end{itemize}

Basically we are using in the protocol both a public key (step
I) and a private key (step II). In fact the message is encrypted (by
A) and decrypted (by B) using the same key,  the sequence of prime
knots that they share (secretly) thanks to  step I.\\

\section{Discussion and conclusions}
There are a number of advantages in basing a cryptosystem
on complex
geometric structures such as knots,
where the selected prime knots could be looked at as
providing  an encryption alphabet.
Note first that the coding procedure that provides the 
Dowker--Thistlethwaite string ({\em e.g.} written in the standard binary
notation) is  efficiently implementable since it grows linearly
with the crossing number. 
As noted above,  
decomposing a composite knot
in its prime components is at least as difficult as
finding the prime factors of a large  number,
while of course the composition (corresponding
to multiplying integers) is an easy task.
In order to attack such kind of protocol, 
one might resort to two strategies.

$\bullet$  The first approach is based on
the use of topological invariants 
which provide, at least in  case of low crossing numbers, 
quite a lot of information. Looking at Knot Tables we note that, up to seven crossings,
all knots are alternating, so that, in particular, the crossing number of
a knot built as a connected sum of alternating knots is the sum of the
individual crossing numbers (but this is not true for non--alternating knots).
On the other hand, most powerful knot polynomials (quoted in  section 2)
are multiplicative with respect to connected sums. So, for instance, we can evaluate \cite{Kau}
the Jones invariant $J_K(t)$ of a given composite knot $K$ getting a (Laurent) polynomial
in the formal variable $t$, but still it is  a hard task  
to extract the polynomials associated with the prime factors of $K$.  
As a matter of fact, such a strategy based on knot invariants 
is effectively unfeasible  
because topological invariants  of polynomial type are not able to
distinguish a (generic) knot from  one of its mutations. 

$\bullet$ Another way is to try to decompose the 
knot diagram containing the message  by resorting
 to iterated combinatorial operations
aimed to recognize in the encrypted message
$ L_{1} \# K'_{1},\, L_{2} \# K'_{2},$
$\, \ldots, L_{N} \# K'_{N}$
at least some of the prime knots
in the public list. But there  exists  no known
algorithm to address the decomposition problem
of a generic knot into its prime components.\\
Finally, as pointed out in section 2, it
is certainly true that 
the recognition problem can be associated
with combinatorially  recursive procedures,
but Haken \cite{Hak} was able to
prove  only the existence of an algorithm running in
exponential time.
On the other hand, the unknotting problem \cite{BiHi}  
--a particular case of the recognition problem
stated in term of comparison of a given knot $K$
with the unknot $K_{\bigcirc}$-- is shown to belong to the complexity 
class $\mathbf{NP}$ \cite{Hass1,Hass2}.\\

The new knot--based cryptographic protocol proposed in this paper
relies on quite simple mathematical notions and 
needs of course to be further specified 
and checked against different types of attacks.
Note however that techniques developed within
the framework of braid group--based cryptography 
(see \cite{BlCiMu}, section 4) do not seem to be implementable
in such a purely topological setting\footnote{
It is worth recalling that the set of (prime) knots equipped with 
connected sum $\#$ is an Abelian semigroup (actually a monoid,
the unknot being the identity element). On the other hand,
closed braids representative of a given knot diagram exist,
but they are certainly not unique. Actually the problem of finding
out the `minimal' braid index for a knot $K$ --namely the minimum number $n$
such that there exists a braid $W$ $\in \mathbf{B}_n$ whose closure reproduces
$K$-- is again a hard problem, {\em cfr.} section 4 of \cite{BiBr}.
Then the knot--based protocol is not effectively reducible
to the group--based approach to classical cryptography. 
}.

In conclusion, it seems quite  promising that,
besides  brute force attacks which would be
exponential resources consuming 
as the topological complexity of the knots grows, more sophisticated
attacks based on (exact or approximate, classical or quantum) 
calculations of polynomial invariants of knots 
are intrinsically unreliable.

\section*{Acknowledgments}

{\small A special debt of gratitude goes to Chiara Macchiavello and Claudio Dappiaggi
for useful conversations.}

%\vfill
%\newpage

\end{document}